# A Utility Use Case: Utilizing Spatiotemporal Data Analytics to Pinpoint Outage Location

Reddy Mandati[1], Po-Chen Chen, Vladyslav Anderson, Bishwa Sapkota, Michael Jarrell Warren, Bobby Besharati, Ankush Agarwal, Samuel Johnston III[2]

[1]Infrastructure and Safety Analytics
Exelon

[2]Regional Distribution Engineering
Baltimore Gas and Electric

*Abstract*—Understanding the exact fault location in the post-event analysis is the key to improving the accuracy of outage management. Unfortunately, the fault location is not generally well documented during the restoration process, creating a big challenge for post-event analysis. By utilizing various data source systems, including outage management system (OMS) data, asset geospatial information system (GIS) data, and vehicle location data, this paper creates a novel method to pinpoint the outage location accurately to create additional insights for distribution operations and performance teams during the post-event analysis.

## I. Introduction

Improving the efficiency of outage management in the distribution grid systems is the one of the top priorities for utilities. Faults in the distribution system can be caused by a variety of factors, including equipment failures and structural damage due to weather-related events. Ideally the exact fault locations could be reported by the crew during the restoration process. However, in the real scenarios, most of the outage locations are not clearly documented or reported under the next upstream operating device which could be several miles away. In this case, in the post-event analysis, finding the exact source and location of fault have been relatively challenging.

There have been many research papers discussing fault location methods including intelligent system methods [1]-[2]; superimposed components methods [3]-[4], impedance based methods [5]–[6], traveling wave methods [7]–[8]; and voltage sag based methods [9]–[10]. While these publications focus on pinpointing the exact fault location by using measurement data, they do not help on eliminating data noise using vehicle data and locate the most accurate fault location.

Big data applications and machine learning methods have been used to address the fault location challenges in the utility industries. The comprehensive literature reviews [11]-[13] show machine learning methods on the fault location. The data analytics method for advanced distribution management system and outage management system are shown in [14]-[15]. Many research articles also reveal the benefits of using smart meter data for fault location applications [16].

Furthermore, to adopt the spatiotemporal nature of various datasets, applications usually require capability of geospatial data processing in order to correlate the datasets on a geographic information systems platform [17]-[18]. While previous studies have approached this problem with different sets of tools and techniques, to the best of authors' knowledge, no study have explored geospatial clustering method to identify the outage location in the post-event analysis so far.

This paper presents a use case that enables the engineers to find potential outage location more efficiently in the post-event analysis. The authors built a novel tool to enable an end-user to evaluate all available data related to an outage to find the actual fault damage location. The more precise fault location data will enable reliability engineering teams to gain insights through trending reliability and asset health concerns at a hyperlocal scale. This information will allow engineers to identify correlation of trouble zones in the same geographic area regardless of whether multiple feeders supply them.

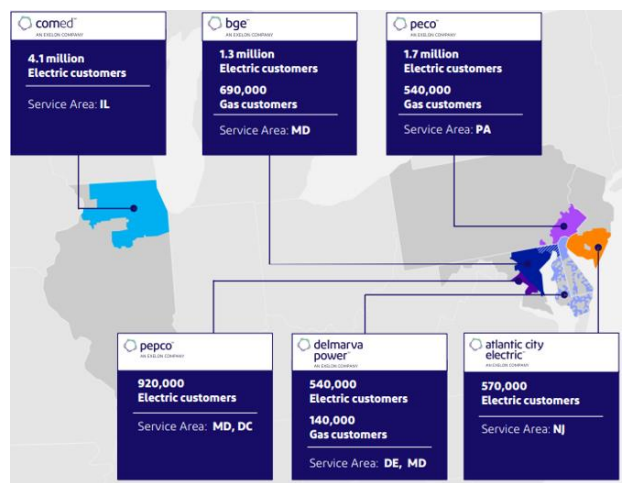

Figure 1. Exelon consists of six regulated utilties.

This work is supported by Exelon and Baltimore Gas and Electric, an Exelon company.

Moreover, outage location is limited to the location of equipment identified in OMS at the time of outage reporting. The accuracy of which equipment is involved in the outage varies due to a variety of factors like the summary of multiple outage calls into a single event (which mostly happens after restoration), incorrect equipment identification due to adverse conditions among other things. Customer location identified via outage is useful for smaller customer count outages but other cases, like a mainline outage, might have so many unique customer calls that customer location value diminishes in relation to being able to identify actual location. In this study, the vehicle location data is applied to solve this issue, and therefore the team can implement more targeted inspection processes and reliability improvement projects generating cost savings for the business.

## II. BACKGROUND

### A. Overview of Exelon

Exelon consists of six regulated utilities and a service company (Fig. 1)[19]:

- Delmarva Power (DPL),
- Potomac Electric Power Company (Pepco),
- Commonwealth Edison (ComEd),
- Baltimore Gas and Electric (BGE),
- Atlantic City Electric (ACE),
- Philadelphia Electric Company (PECO),
- Business Service Company (BSC).

Exelon's utilities deliver electricity to approximately 10 million customers in Delaware (Delmarva Power), the District of Columbia (Pepco), northern Illinois (ComEd), Maryland (BGE, Delmarva and Pepco), New Jersey (Atlantic City Electric) and southeastern Pennsylvania (PECO) [19].

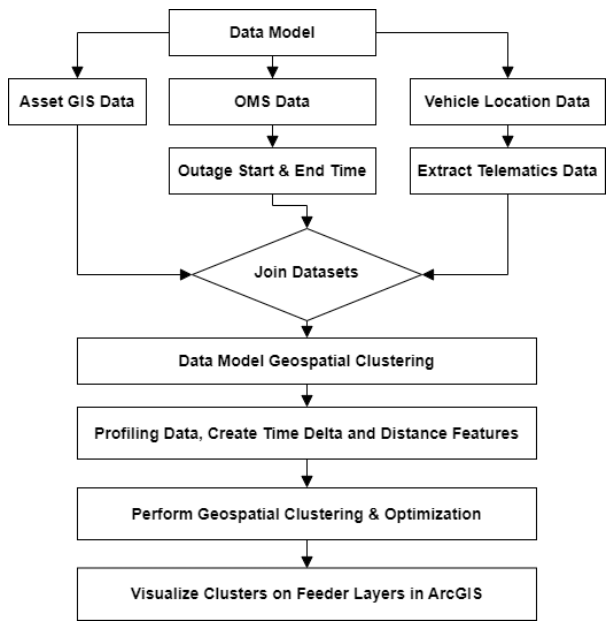

Figure 2. Overall methodology.

### B. Use Case Background and Formulation

The goal is to integrate outage location, electric asset GIS location, and vehicle location information to enhance predictability to determine true fault/outage location. The formulation of use case is shown below.

1. Current state assessment: understand existed engineering investigations process to identify analytic opportunities.
2. Stakeholder onboarding: review opportunities and confirm value in developing analytics, and building-out network to oversee, champion and drive adoption.
3. Data collection and preparation: extract input data and perform data cleansing, and then build the data model.
4. Execution and modeling: develop analytics using agile approach with regular check-ins with stakeholder network.
5. Validate results with subject matter experts: validate the results with the stakeholder and promote the tool for wider usage.

## III. METHODOLOGY

The overall methodology is shown in Fig. 2. After the datasets are investigated and processed, they are used to develop the model. Then the geospatial clustering and optimization techniques are applied to pinpoint the fault location. Some of the data investigations attempt to answer the following comments regarding improving fault location accuracy:

- Correlations between feeder properties (e.g. overhead, underground, exposure & length, 13 kV vs 34 kV, metro vs suburban vs rural) and corresponding outage data to yield descriptive insights,
- Word use patterns and asset references within the outage comments,
- Human performance trends (accuracy and completeness of entries) can be used to identify, predict, or exclude certain outage types (e.g. if outage comment entries for certain outage types are more or less completed and accurate than others),
- Correlation between certain vehicle types and outage cause or location (e.g. if certain vehicle types and corresponding movements provide more value than others).

### A. Data Cleansing and Preparation

Quality of outage and crew location data are critical to the accuracy and usefulness of insights. During the initial investigation, we discovered many scenarios where the data may generate irrelevant insights – though they represent just a fraction of the total events.

There are three sources of data used in this use case:

- OMS Data: historical outages, cause, affected customers, activities of crew on field, outage restoration, outage duration and performance of feeders. Note the OMS data already accounts for AMS data interruption signal.
- Asset GIS Data: tripped equipment location, active status, bounding boxes around feeder to filter assets data spatially. The purpose of using GIS in this analysis is to

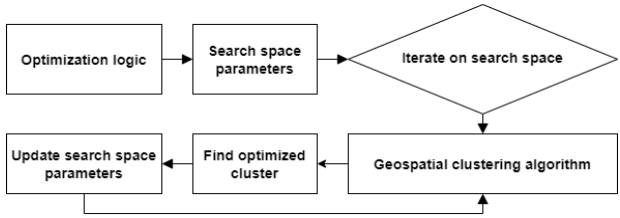

Figure 3. Optimization Logic

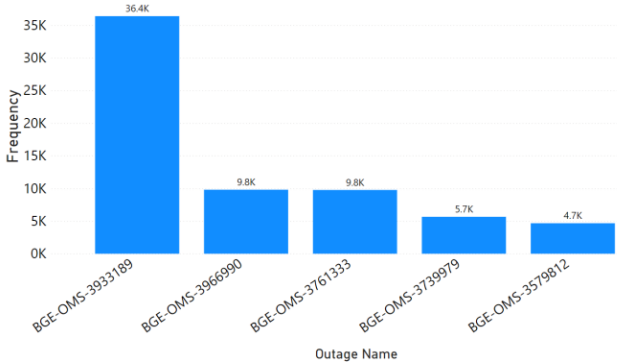

Figure 4. Top 5 outages with frequency of vehicle points.

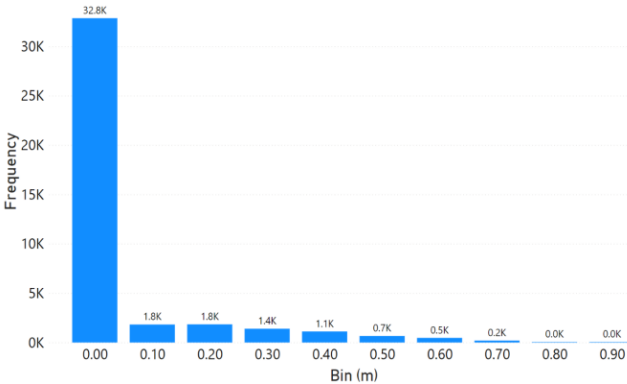

Figure 5. Distance between vehicle points in meters (0-1000) for Sample Outage.

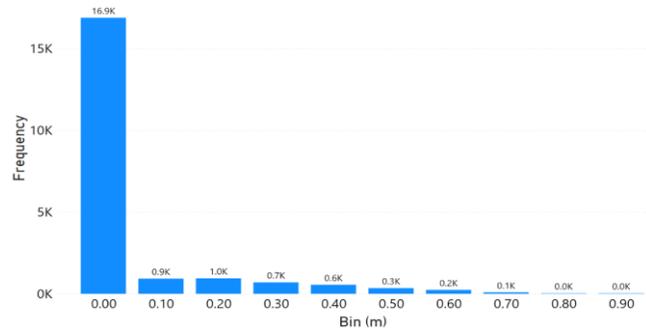

Figure 6. Top 5 Distribution of vehicle points in meters for outages.

- filter vehicle data spatially alongside time-based filtering using outage information.
- Vehicle location data: Points in time where work vehicles are stationary near an OMS identified outage location and time of outage. The stationary vehicle locations in time are our novel data source in our attempt to get an outage location more accurately than the initially identified location from first responders. Telematics data from crew vehicles enables us to decide where repair was made.

Once the data correlation model is created for predictive analytics, the visualization is created on ESRI platform [20]. To find where the crew spent most time when the outage occurred, the geospatial clustering on crew vehicle coordinates is performed. The time delta and distance features, such as duration and distance between crew vehicle location, is calculated every time crew vehicle transmits signal. This provides a baseline for configuring hyper parameters in geospatial clustering. The two main hyper parameters needed are distance between vehicle points and number of vehicle points associated with outage- we refer these two hyper parameters as search space parameters in optimization logics shown in Fig. 3.

### B. DBSCAN Geospatial Clustering

Clustering is the process of grouping large data points in more than one group based on their similarity with neighboring data points. Density Based Spatial Clustering of Applications with Noise (DBSCAN) is one of the spatial clustering techniques to separate clusters of high density from clusters of low density [21]. DBSCAN shows good performance in data points which have high density observations compared to the data points that are not very dense with observations. In this study, vehicle points threshold for each outage for dense vehicle points were determined by visually inspecting outage layers.

The key points in DBSCAN are the core points and the reachability combined to generate clusters and filter out noise. In this case the optimization logic is developed (Fig 3). To better understand the characteristics of crew vehicles from the initial clusters, the following data attributes were considered. These attributes contribute to the understanding of the targeted crew vehicles assigned to specific outage.

- The number of unique crew vehicles in the cluster,
- The time each crew vehicle spent at the cluster location,
- The order of vehicle arriving/leaving at the cluster location.

The process satisfies two properties: 1) all points within the cluster are mutually density-connected, 2) if a point is density-reachable from some point of the cluster, it is part of cluster as well. In DBSCAN, the quality of cluster depends on distance measure. It is vital to understand distribution of vehicle points per outage (parameter 1) and distance between vehicle points in time (parameter 2) to come up with search space parameters. Values of search space are drawn randomly to iterate through geospatial clustering algorithm and find optimized cluster with confidence scores and update search space parameters based on high confidence scores.

Determining the optimal distance(parameter 2) was challenging due to varying density of vehicles across different outages. We employed a k-distance graph analysis to identify

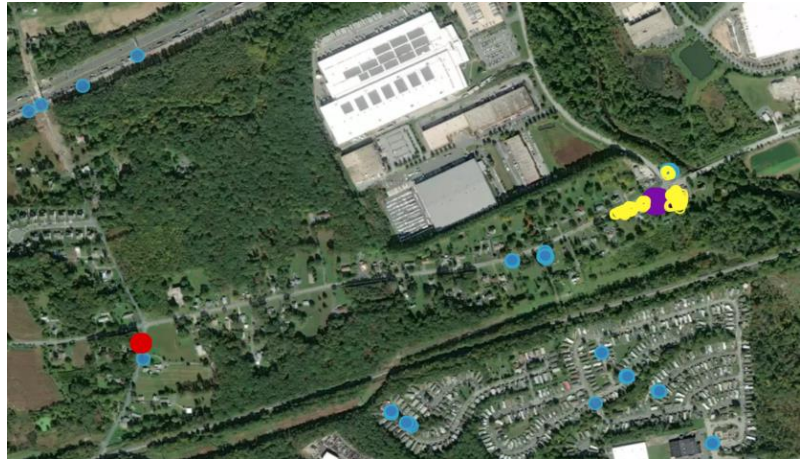

(a) Initial clustering data with noise data.

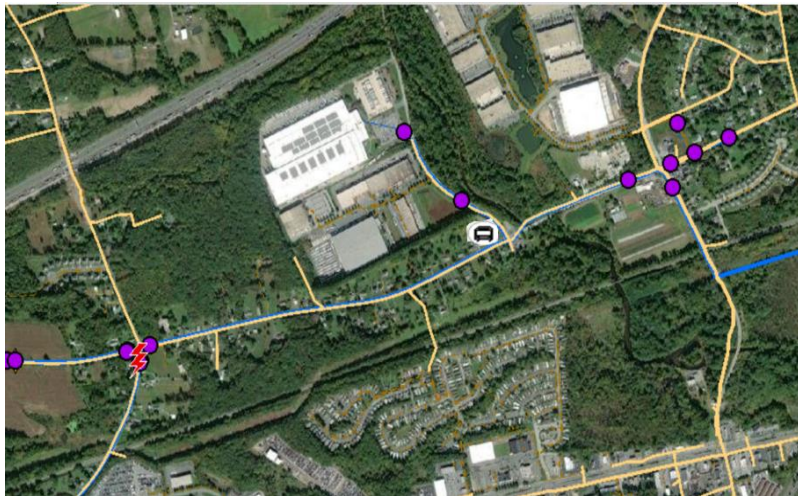

(b) Final clustering data, where vehicle centroid location represents the predicted outage location, and the noise points are removed.

the elbow point that best represents the transition from dense to sparse regions. Additionally, we conducted sensitivity analysis to understand the impact of minimum vehicle points per outage (parameter 1) on cluster formation, ensuring robustness across diverse outage scenarios.

Fig. 4 shows top few sample outages with their frequency of vehicle points. For the first outage, there are approximately 36,000 vehicle points which have unique crew vehicle data points associated within outage duration. Clustering works effectively when we have outages with more crew vehicle points, however, this does not apply for few outages with less vehicle points. This distribution of outages and their frequency will help in determining threshold for search space parameters in optimization logic Fig3.

Fig. 5 shows top few sample outages with distance between vehicle points in time and their frequency, the range is 0 to 1000 meters for all vehicles, this distribution will help in determining threshold for search space parameters in optimization logic as this is one of the hyper parameters in geospatial clustering.

Fig 6 shows distribution of distance between vehicles and frequency of crew vehicles used in search space parameters from a sample outage. If the distance between vehicle points is closer the cluster formed will be dense and more accurate. If the distance between vehicle points is wide (e.g. 500-1000 m) in the search space, the inaccurate cluster will be formed and not remove any noise (all vehicle points will be part of cluster- which is false).

IV. RESULTS

A. Example

Fig. 7 (a) shows a sample outage with crew vehicles overlaid on the map. It is imperative to eliminate noise to improve determination of accurate fault location and hence improve the accuracy. The blue points are considered noise whereas the points in yellow are our area of interest. Since the reported outage location (the red point) in OMS data is not reliable, the idea is to identify the more reliable outage location using geospatial techniques. For this, all the vehicle telematics data corresponding to this outage were fetched and the initial geospatial clustering was performed. The vehicle data considered to be the noise (the blue points) are then eliminated from the study. This leaves the yellow points which are the locations of interest. It should be noted that while there are many vehicle points in the proximity of the reported outage location area, only the yellow ones are selected by the clustering algorithm as non-noise.

With the derivation of predicted outage locations, visualization process was carried out. First, several layers including electric assets (isolating, wire, cable) were fused with outage location layers that was predicted earlier. The ESRI platform was used to create dashboard to render these layers and perform necessary queries (Fig. 7(b)). The dashboard was provided to business stakeholders to perform quality assessment and receive feedback for revisions. Currently, out of 232 outages verified, 180 of them, 78%, are accurate predicted fault location from clustering model side. We are still in the process of tuning the models and look at better ways of reducing noise from inaccurate input data.

*B. Deployed Fault Location Analytics Tool*

Deployed Fault Location Analytics Tool displays not only the pinpointed outage location, but also other assets linked to outages, including feeder, overhead and underground cables, isolating equipment, and substation locations. The information extracted from NLP text analytics and geocoding of converting text to address from outage data, where crew logs in their findings after inspection of asset, is critical. In addition, adding in additional meta data related to outage will enable users to understand various device types and its behavior for outages.

## V. Conclusion

The contributions are summarized below:

- Propose a novel geospatial clustering framework to pinpoint the fault location of an outage based on outage data, asset GIS data, and crew vehicle telematics data: While the datasets used in this paper are from BGE, the data science development cycle followed may ensure the proposed application is scalable to other utilities.

- Demonstrate how data processing and cleansing are performed and how the insights are observed from the correlation of various datasets.

- Developed DBSCAN and optimization algorithms to identify the fault location based on the clustering results.

- Deployed a web application tool for stakeholder to use.

Building and deploying a scalable optimized geospatial clustering tool is currently in progress for various sizes of feeders in BGE. The biggest technical challenge lies in building data-centric geospatial algorithms as they are constantly searching for best parameters to create an optimized cluster, which can be computationally expensive, leveraging cloud based distributed computing can reduce model development time and future work will focus on algorithm optimization to enhance scalability.

Future research will explore scaling the geospatial clustering modeling to other Opcos with in Exelon to extend the capabilities of pin-pointing fault location.

Acknowledgment

The authors gratefully acknowledge the contributions of Catherine Jeffcoat, BGE Regional Distribution Engineering, for the innovative project ideas and support. The authors would like to also thank Amin Tayyebi, Ratish Kutty, Ryan McGuire, and Summerlyn Turner for their contributions on this work.